\documentclass[11pt]{article}
\pdfoutput=1
\usepackage[utf8x]{inputenc}
\usepackage{amssymb,amsmath,mathrsfs,enumerate}
\usepackage{graphicx,rotate,multicol,braket, multirow}
\usepackage{url}
\usepackage{cite}
\usepackage{soul}
\usepackage{braket}
\usepackage[normalem]{ulem}
\usepackage{wrapfig}
\usepackage[textfont=it,labelfont=bf]{caption}
\usepackage{bm}
\usepackage[colorlinks=true,
linkcolor=red,
urlcolor=blue,
citecolor=blue]{hyperref}
\usepackage{lineno}
\usepackage{color}
\usepackage{bbold}
\usepackage[dvipsnames]{xcolor}
\long\def\rpl#1!!#2!!{\textcolor{red}{#1} \textcolor{blue}{#2}}

\usepackage[T1]{fontenc}

\newcommand{\be}{\begin{equation}}
\newcommand{\ee}{\end{equation}}
\newcommand{\bea}{\begin{eqnarray}}
\newcommand{\eea}{\end{eqnarray}}

\evensidemargin=\oddsidemargin
\newcommand{\RomanNumeralCaps}[1]
    {\MakeUppercase{\romannumeral #1}}

\def\Eqn#1{Eq.\ (\ref{#1})}

\allowdisplaybreaks

\title{\Large\bf 
Islands in Bianchi type $\rm \RomanNumeralCaps{1}$ universe
}

\author{
  \sf 
  Ido Ben-Dayan,$^{a,}$\footnote{ido.bendayan@gmail.com}
\quad  Merav Hadad,$^{b,}$\footnote{meravha@openu.ac.il}
\quad Ayushi Srivastava$^{a,}$\footnote{srivastavaayushi860@gmail.com}
\\[10pt]
    \small\em$^a$Physics Department, Ariel University, Ariel 40700, Israel\\
    \small\em $^b$Department of Natural Sciences, The Open University of Israel, Raanana 43107, Israel
\\  }

\date{}

\begin{document}


\maketitle

\begin{abstract}
We study the conditions for finding an island in an anisotropic universe---Bianchi type $\rm \RomanNumeralCaps{1}$ filled with radiation. We verify that the existence of islands does not depend on their shape. We then find that islands may form at certain times, near the turnaround point---where the universe turns from contraction to expansion in one of the directions.
This is in line with previous analyses regarding cosmological space-times where islands form if one has two energy scales in the problem, such as the typical temperature of the universe and, on top of that, cosmological constant, curvature, anisotropy, or some other mass scale.
\end{abstract}

\bigskip

\section{Introduction} \label{s:intro}

The black hole information paradox is a long-standing theoretical physics puzzle arising from the apparent conflict between quantum mechanics and general relativity. Hawking\cite{hawking1975particle,hawking1976breakdown} applied the semiclassical quantum field theory approach in curved space-time and revealed that black holes emit radiation, known as Hawking radiation. The calculation suggested that the final state of radiation would retain information only about the initial state's total mass, electric charge and angular momentum, implying information loss. This conflicts with the principle of unitarity in quantum mechanics, which asserts that a pure state should evolve into a pure state.

As the black hole evaporates, its area---and, consequently, the Bekenstein-Hawking entropy---decreases. Regarding the radiation, initially its entropy is zero. As the black hole evaporates, the fine-grained entropy of the Hawking radiation increases without bound.

Page\cite{page1993information,page1993average} later proposed that for the process to be unitary, the entropy of the radiation should follow the Page curve, initially increasing and then decreasing back to zero as the black hole evaporates, indicating a return to a pure state. Recent developments in semiclassical gravity have demonstrated that the Page curve can be reproduced by using the quantum extremal surface (QES) prescription for computing the entropy of the radiation\cite{ryu2006aspects,ryu2006holographic,hubeny2007covariant,faulkner2013quantum,engelhardt2015quantum}. After the Page time, when the black hole radiated away half of its entropy, an entanglement island\cite{almheiri2019entropy,almheiri2020page,almheiri2020replica,penington2020entanglement,penington2022replica,lewkowycz2013generalized}\footnote{These models do not have an explicit construction of the replica wormhole. An explicit construction of these is followed in higher-dimensional holographic models\cite{geng2024replica}. }---a disconnected region from the reference system---appears. The entropy of a nongravitational system $R$ is computed using the island formula, a generalized version of the Ryu-Takayanagi (RT) formula for holographic entanglement entropy. It is expressed as,
\begin{eqnarray}
\label{e:islandformula}
S(\textbf{R})=\underset{I}{min}\,\,  \underset{I}{ext}\left(S_{\rm gen}(R \cup I) \right)  = \underset{I}{min} \left\{ \underset{I}{ext}\Big[\frac{A(\partial I)}{4G_N} + S_{\rm matter}(R \cup I)   \Big] \right\}\,.
\end{eqnarray}
The first term, known as the ``area term," resembles the Bekenstein-Hawking entropy associated with black hole horizons. Here, $A(\partial I)$ represents the area of the boundary of the island, and $G_N$ is the gravitational constant. The second term, referred to as the ``matter term", $S_{matter}(R\cup I)$, represents the quantum entanglement entropy of the bulk fields across the combined region $R\cup I$.
The $S(\textbf{R})$ with boldface $\textbf{R}$ signifies the entropy computed by the QES
prescription, and the use of a nonboldface $R$ indicates the appearance of von Neumann entropy computed directly from the semiclassical state. The generalized entropy is extremized (ext) over the choice of $I$. If multiple extremal surfaces exist, the one that minimizes (min) the right-hand side of the equation is chosen, as discussed in \cite{faulkner2013quantum}. 
\begin{figure}
	\centering
	\includegraphics[scale=0.15]{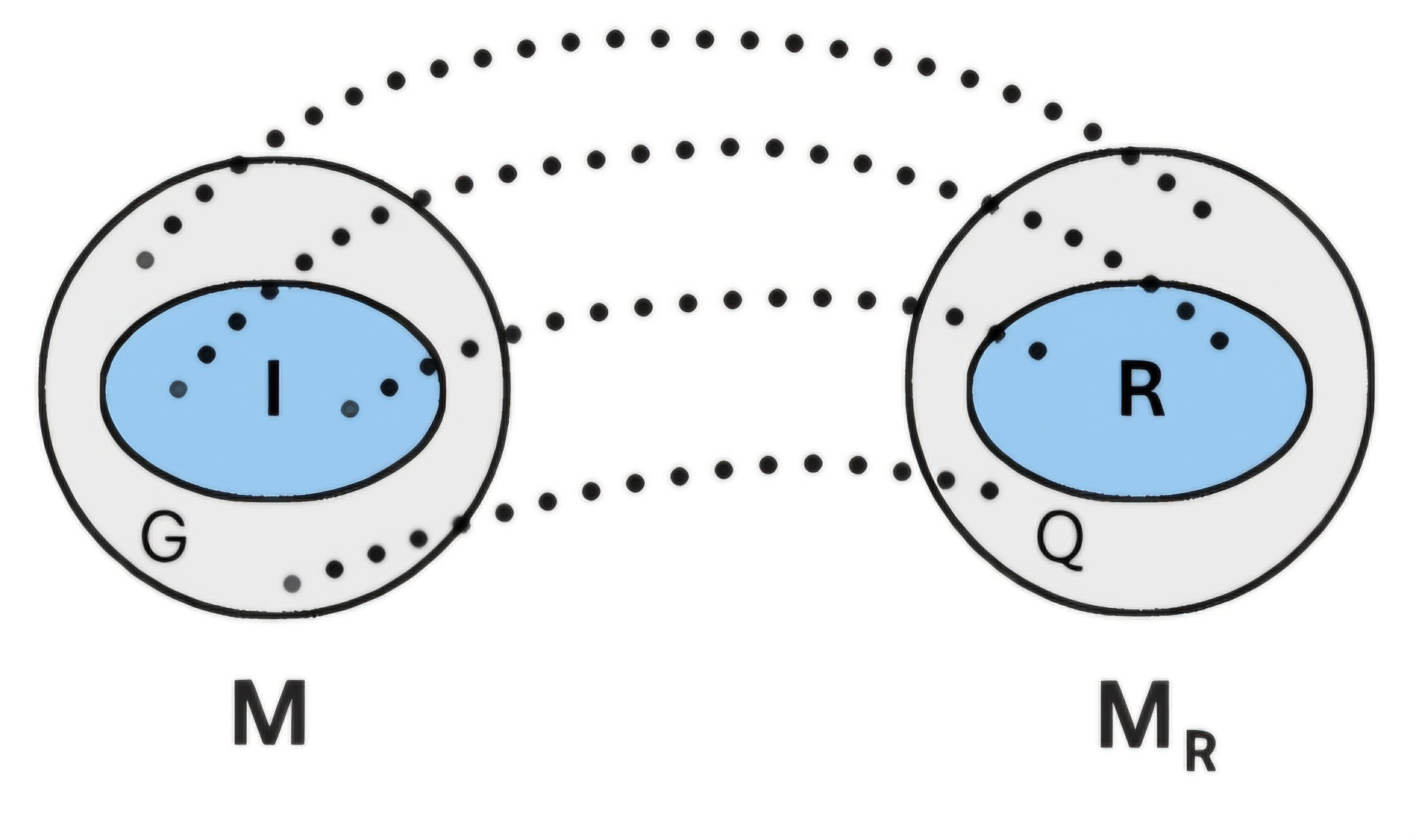}
	\caption{\em\small Schematic figure of $M$ and $M_R$. A region in $M$ can be an island $I$ for some reference region $R$ only if it satisfies the necessary conditions. The entanglement between the two regions is shown by the dotted lines. $G$ is the complement of $I$, and $Q$ is the complement of $R$.
		\label{fig:v5}}
\end{figure}
In the black hole context, $R$ is the region outside the black hole where  Hawking radiation is obtained (or expected), $I$ is (mostly)
inside the black hole, and they are related by the large entanglement between the interior and
exterior Hawking pairs. Right after the black hole forms, no Hawking radiation has escaped yet. There is no extremal surface inside the black hole. So, initially, the entropy of radiation is $S(\textbf{R})\approx S_{\rm matter}(R)$. If the black hole starts in a pure state, the initial entropy is zero. As the black hole starts emitting Hawking radiation, the von Neumann entropy of the radiation grows linearly as more and more Hawking quanta are emitted. As the evaporation proceeds, a nontrivial island appears in the inside of the black hole. Its boundary is very close to the black hole horizon and extends almost through the whole interior. Consequently, the partners of Hawking quanta that fall inside the black hole are almost contained in the island. This reduces the contribution of the matter term. The dominant contribution now comes from the area term. As the black hole horizon shrinks, the area term decreases and finally vanishes. Thus, the Page curve
is followed, and the black hole evaporation process is unitary.

Associated with island formation is a violation of the Bekenstein area bound, which states that the fine-grained entropy of a region  $S$ is bounded by the Bekenstein–Hawking entropy of the black hole \cite{bekenstein1981universal},
\begin{eqnarray}
\label{e:areabound}
S \leq \frac{\rm Area}{4G_N}\,.
\end{eqnarray}
Interestingly, the Bekenstein bound is also violated in cosmology.
Hartmann \textit{et al.} \cite{hartman2020islands} then posed the question of whether such a violation can lead to the appearance of an island in our Universe. If it could, then what would be the analog of the Hawking radiation that might produce the large amount of entanglement entropy necessary for the formation of islands. They considered favorable assumptions about the entanglement structure that favor the existence of islands. They examined the radiation-dominated, flat Friedmann-Lema\^{i}tre-Robertson-Walker (FLRW) space-time $M$ with zero, a positive and negative cosmological constant, which is in an entangled state with a nongravitating system $M_R$, see Fig.\ref{fig:v5}. They concluded that islands appear only in the case of a negative cosmological constant. They reached this conclusion by giving three conditions nearly sufficient to identify the islands in a given space-time. 

Here, we rewrite these conditions for a nonempty island in space-time $M$ entangled with the reference system $M_R$, in a thermofield-double-like state as introduced in \cite{hartman2020islands}. These conditions are given as follows: 

\textit{Condition 1:} The Bekenstein area bound must be violated by the island region
\begin{eqnarray}
	\label{e:cond1}
	S(I)> \frac{A(\partial I)}{4 G_N}\,,
\end{eqnarray}
where $S(I)$ is the matter entropy of the island. We denote such an island candidate as $I$ and its complement on its Cauchy slice as $G$.

\textit{Condition 2}: The formation of an island is governed by the condition that the generalized entropy, $S_{\rm gen}(R \cup I)$ is extremized. Specifically, if the island has a boundary, the derivative of the generalized entropy along the null direction must be zero. This implies the following bound:
\begin{eqnarray}
	\label{e:sgenb1}
	\pm \frac{d}{d\lambda_\pm}S_{\rm gen}(I) \geq 0\,.
\end{eqnarray}
The above inequality means that the boundary of an island must be a quantum normal region; that is, the region where the quantum expansion is positive in an outward direction and negative in an inward direction.
Here the generalized entropy, $S_{\rm gen}$, is defined as 
\begin{eqnarray}
	\label{e:sgeni}
	S_{\rm gen}(I) = \frac{A(\partial I)}{4G_N} + S(I) \,.
\end{eqnarray}
The derivatives are null deformations of the
boundary of the island, with $d/d\lambda_+$ outgoing and $d/d\lambda_-$ ingoing deformation with respect to $I$. Suppose $k^\mu = \partial x^\mu / \partial \lambda $ is the null vector field normal to the surface, where the surface is specified by setting a single function, say $f$, to a constant and $x^\mu(\lambda)$ are the coordinates. 
Then, the vector field
\begin{eqnarray}
\label{e:kmu2}
k^\mu=g^{\mu \nu}\nabla_\nu f\,,
\end{eqnarray}
will be normal to the surface. For it to be a null vector field, we will impose the condition that $k_\mu k^\mu=0$. Then $d S_{\rm gen}/d\lambda = k^\mu \partial_\mu S_{\rm gen}$.

\textit{Condition 3}: Let $G$ be a region that surrounds the island and shares a boundary, and we assume that it is spacelike separated from region $R$. Then, the third condition states that
\begin{eqnarray}
\label{e:thirdcond}
\pm \frac{d}{d\lambda_\pm}S_{\rm gen}(I) \leq \pm \frac{d}{d\lambda_\pm} [S_{\rm matter}(I)-S_{\rm matter}(G)]\,.
\end{eqnarray}
It provides an upper bound on $\pm d S_{\rm gen}(I)/{d\lambda_\pm}$. Since $I$ and $G$ share a boundary, it is only along this shared boundary that the deformation affects that area term. The above equation can be reinterpreted as $G$ being a quantum normal region
\begin{eqnarray}
	\label{e:sgenb2}
	\pm \frac{d}{d\lambda_\pm} S_{\rm gen}(G) \leq 0 \,.
\end{eqnarray}
The deformation $d/d\lambda_+$ is ingoing with respect to $G$. The generalized entropy for the region $G$ is defined as 
\begin{eqnarray}
	\label{e:sgen2}
	S_{\rm gen}(G) = constant-  S(I) + \frac{A(\partial I)}{4G_N} \,,
\end{eqnarray}
from $S(G)=constant-S(I)$ \cite{hartman2020islands}. The constant in $S_{\rm gen}(G)$ is the matter entropy of the entire slice containing $I$ and $G$,
but this constant will drop out of the conditions. Taken together, these three conditions are very restrictive. So, if they are all satisfied, there is a strong hint that an island can be found.

The work in \cite{hartman2020islands, bousso2022islands, espindola2022islands} explored the problem of entanglement islands\footnote{The work mentioned here focuses on specific models. The explicit construction of islands in cosmology in general dimensions is done in \cite{Geng:2021wcq}.} in radiation-filled FLRW universes, examining the impact of spatial curvature and the cosmological constant on their existence. The analyses showed that in the presence of curvature, islands can exist near the ``turnaround'' time when the universe switches from expansion to contraction or vice versa. In \cite{ben2023islands}, results were generalized, finding that islands are relatively common in cosmology provided a certain perfect fluid equation of state. Moreover, these islands are not necessarily near the turnaround time. Thus, if there exist two energy scales, such as the temperature of radiation and curvature, islands tend to form, while if there is only a single energy scale, such as the temperature of radiation, no islands form. 
In this manuscript, we present our study on the effect of anisotropy on the existence of an island, again consisting of an additional energy scale, that of anisotropy. We investigated a Bianchi type $\rm \RomanNumeralCaps{1}$ model of the universe, which is flat with radiation domination. Since the universe is anisotropic, different spatial directions can contract or expand at different times.
Our findings indicate that entanglement islands can indeed exist in this anisotropic toy model within the semiclassical regime. We find the island forms again near a ``bouncing behavior" 
when the scale factor(s) switches from contraction to expansion in one or more directions. 

In Sec. \ref{s:flrw_rad}, we will examine FLRW models filled with radiation with zero curvature and zero cosmological constant. We will summarize the results of previous work done on the spherical island case and extend this analysis to the ellipsoidal island. This serves as a check that the possible existence of an island does not depend on its shape. In Sec. \ref{s:model}, we present a detailed study of the Bianchi type $\rm \RomanNumeralCaps{1}$ model filled with radiation. We study the conditions for the existence of a spherical or ellipsoidal island in this model. We found that islands can exist for $t\ll t_0$, where $t_0$ is the time where the BI model is momentarily FLRW model, and all scale factors exponentially approach the same value. The formation occurs at time $t$ when one or more of the scale factors show the bouncing behavior, i.e., they switch from initial contraction to expansion. Some technical results were relegated to the Appendix.
\section{FLRW universe filled with radiation}
\label{s:flrw_rad}
Consider the FLRW universe filled with radiation with an equation of state $w=1/3$. The metric for this flat universe is
\begin{eqnarray}
\label{e:flrwmetric}
\mathrm{d}s^2=-\mathrm{d}t^2+a(t)^2\left[\mathrm{d}\chi^2+\chi^2 \mathrm{d}\Omega^2\right]=a^2(\eta)\left[-\mathrm{d}          \eta^2+d\chi^2+\chi^2 d\Omega^2\right]\,,
\end{eqnarray}
where $t$ denotes cosmic time and $\eta$ conformal time, which are related by $dt=a(\eta)d\eta$. Here, $a$ is the scale factor, which is a dimensionless quantity and is given as 
\begin{eqnarray}
\label{e:scalefac}
a=a_0\sqrt{\frac{t}{t_0}}\,.
\end{eqnarray}
Here, the subscript ``$0$'' denotes some time of normalization. In this parametrization, the big bang singularity is at $a=0$. 
The governing equations of motion are given by the first Friedmann equation and the continuity equation,
\begin{eqnarray}
H^2&=&\frac{8\pi G_{N}}{3}\rho_{\rm rad}\,, \label{e:Friedmann1}\\
\frac{d\rho_{\rm rad}}{dt}&=&-3H(\rho_{\rm rad}+p_{\rm rad})~,\label{eq:continuity_i}
\end{eqnarray}
where $\rho_{\rm rad}$ is the energy density, $p_{\rm rad}=1/3\, \rho_{\rm rad} $ is the pressure density, $t$ is time and $H$ is the Hubble parameter.  The energy density of radiation behaves as
\begin{eqnarray}
\label{e:rhor}
\rho_{\rm rad} = \rho_{0}\left(\frac{a}{a_0}\right)^{-4} \,,
\end{eqnarray}
and it is conformally related to a finite-temperature state in Minkowski space-time as
\begin{eqnarray}
\label{e:cftcth}
\rho_0 = c_{\rm th}T_0^4\,.
\end{eqnarray}
Here, $c_{\rm th}=\pi^2 g_*(T)/30$ is a constant and $g_*(T)$ counts the total number of effective massless degrees of freedom. Assuming the Standard Model (SM), its value for different temperature ranges is given by
\begin{eqnarray}
\label{e:g*}
g_*(T\ll {\rm 1 \,\,Mev})&=& 3.36\,,\cr
g_*(1\,\,{\rm Mev}<T<100\,\, {\rm Mev})&=&10.75\,,\cr
g_*(T>300\,\,{\rm Gev})&=&106.75\,.
\end{eqnarray}
For extensions of the SM, for example, in string theory, it can receive higher values even by an order of magnitude or more. Time is related to energy density and temperature via the following relation:
\begin{eqnarray}
\label{e:time}
\frac{1}{4t^2}=\frac{8}{3}\pi G_N \rho \implies t&=&\frac{1}{2\sqrt{\frac{8}{3}\pi G_N \rho_{rad} }}\,=
\frac{1}{2\sqrt{\frac{8}{3}\pi G_N c_{\rm th}T^4 }}\,.
\end{eqnarray}
In local thermal equilibrium, the comoving entropy density $s_c= s a^3$  of the system is constant in time. Here, $s$ is the thermal entropy density. Thus, the entropy of a domain $S=s \rm{V}=s_c \mathcal{V}$, where $\rm{V}$ is the physical volume and $\mathcal{V}$ is the coordinate volume.
At the time of normalization $t_0$, the comoving entropy is
\begin{eqnarray}
\label{eq:flrwsc}
s_c=\frac{4\rho_0a_0^3}{3T_0}=\frac{4 c_{\rm th}T_0^3a_0^3}{3} \,.
\end{eqnarray}
The constancy of the comoving entropy is essential for the conclusions we shall derive here. 
We are interested in islands away from the singularity where semiclassical analysis can be trusted, that is,  $\rho_0 \ll (8\pi G_{N})^{-2}$ and $t_0>>\sqrt{6\pi G_N}$. We will consider the case of spherical and ellipsoidal islands candidates. We show that even though there is a region where all the conditions are satisfied, it leads to an island near the FLRW singularity. It is a minimum of the generalized entropy in the time direction, so it entails a formal violation of quantum focusing \cite{Bousso:2015mna,Shahbazi-Moghaddam:2022hbw,Ben-Dayan:2023inz}. However, this is outside the validity of the semiclassical theory\cite{hartman2020islands}.
\subsection{Spherical island candidate}
Consider a spherical island of radius $\chi$ in the FLRW radiation-dominated universe with the metric as defined in \Eqn{e:flrwmetric}. 
The conditions necessary for the existence of the island at time $t$ are found as \cite{ben2023islands},
\begin{eqnarray}
\label{e:flrweqnn}
\mbox{\textit{Condition 1:}}\quad   s_c&>&\frac{a^2(t)}{4G_N}\frac{\mathcal{A}(\chi)}{\mathcal{V}(\chi)}\,,
\\
\mbox{\text{Condition 2 and 3:}}\quad  s_c&>&\frac{a^2(t)}{4G_N} \left( 2 a(t) |H|\frac{\mathcal{A}(\chi)}{\mathcal{V}^\prime (\chi)} + \frac{|\mathcal{A}^\prime (\chi)|}{ \mathcal{V}^\prime (\chi)}   \right)\,.
\end{eqnarray}
After substituting \Eqn{eq:flrwsc} and the area and volume of the sphere in the above equation, we get the following:
\begin{eqnarray}
\label{e:con1.5}
\mbox{\text{Condition 1:}}\quad  G_N \rho_{0} &>& \frac{9}{16}\frac{T_0}{\chi_{\rm Phys}}\left(\frac{a}{a_0}\right)^3\,,
\\\label{e:con23.5}
\mbox{\text{Condition 2 and 3:}}\quad G_N \rho_{0} &>& \frac{3T_0}{8}\left( \frac{a}{a_0}\right)^3 \left[ |H| + \frac{1}{\chi_{\rm Phys}} \right]\,,
\end{eqnarray}
where $\chi_{\rm phys} \equiv a \chi$ is the physical radius of the island candidate. Now $\chi$ can be taken as large as required such that condition $1$ can always be satisfied. Conditions $2$ and $3$ after substituting \Eqn{e:Friedmann1} become
\begin{eqnarray}
\label{e:con2.2}
	\mbox{\text{Conditions 2 and 3:}} \quad G_N \rho_{0} &>& \frac{3T_0}{8}\left( \frac{a}{a_0}\right)^3 \left[ \sqrt{\frac{8\pi G_N \rho_0}{3}}\left(\frac{a}{a_0}\right)^{-2} + \frac{1}{a \chi} \right]\,.
\end{eqnarray}
In the limit $\chi^{-1}<< (G_N \rho_0)^{1/2} $, the second term in the above equation can be neglected and substituting $\rho_0 = c_{\rm th}T_0^4$ for radiation, we get 
\begin{eqnarray}
\label{e:con1.3}
T>\sqrt{\frac{3\pi }{ 8 G_N c_{\rm th}}}\quad \Leftrightarrow \quad t< \sqrt{\frac{2 G_N c_{\rm th}}{3\pi^3}}\,,
\end{eqnarray}
where from the Friedmann equation, we have used the following relation between $T_0$ and $t_0$, 
\begin{eqnarray}
	\label{e:Tt0rel}
	t_0=\frac{1}{\sqrt{\frac{32}{3} \pi G_N c_{\rm th}}T_0^2}\,.
\end{eqnarray}
To stay in the semiclassical regime, we want $t>>(6 \pi G_N)^{1/2}$. This means that \Eqn{e:con1.3} will not be satisfied unless $c_{\rm th}$ is very large. For the SM in the early Universe with $g_*(T > 300\,\, \rm GeV)=106.75$, it is still not enough. In conclusion, the above equation implies that an island does not exist in a radiation-filled FLRW universe, and it also cannot evolve into a Universe with an island because of the conformal relation between $\rho_0$ and $T_0$, in accord with \cite{hartman2020islands, bousso2022islands, espindola2022islands, ben2023islands}.
However, in certain extensions of the SM, like string theory, a multitude of light particles, such as axions, are assumed, resulting in $c_{th} \gg 1$. Hence, if one of these models is realized in Nature, one can expect an island in the early Universe still in the semiclassical regime.
\subsection{Ellipsoidal island candidate}
Consider the following metric of the flat FLRW model in Cartesian coordinates:
\begin{eqnarray}
\label{e:flrwellipmetric}
\mathrm{d}s^2=-\mathrm{d}t^2+a_f(t)^2\left[\mathrm{d}x^2+\mathrm{d}y^2+ \mathrm{d}z^2\right]\,.
\end{eqnarray}
Let us examine the case of an ellipsoidal island to verify that the existence of an island is independent of its shape. The equation of an ellipsoid is given by
\begin{eqnarray}
\label{e:sufofell}
\frac{x^2}{a^2}+\frac{y^2}{b^2}+\frac{z^2}{c^2}=r^2\,.
\end{eqnarray}
The above equation of an ellipsoid with axes $a$, $b$, and $c$ can be satisfied using the following parametrization: 
\begin{eqnarray}
\label{e:flrwellippara}
x&=& a\, r \sin\theta \cos\phi\,,\nonumber\\
y&=& b\, r \sin\theta \sin\phi\,,\nonumber\\
z&=& c\, r \cos\theta\,,
\end{eqnarray}
where $r$, $\theta$, and $\phi$ are defined as
\begin{eqnarray}
\label{e:randtheta}
r=\sqrt{\frac{x^2}{a^2}+\frac{y^2}{b^2}+\frac{z^2}{c^2}}\,, \quad \tan\theta = \frac{\sqrt{\frac{x^2}{a^2} + \frac{y^2}{b^2}}}{\frac{z}{c}}\,,\quad \tan\phi=\frac{a y }{b x}\,.
\end{eqnarray}
Here, $r$ is a scaling factor that varies from $0$ to $1$, while $\theta$ and $\phi$ range from $0$ to $\pi$ and $0$ to $2\pi$, respectively. 
The volume and surface area of the ellipsoid are 
\begin{eqnarray}
\label{e:eareaa}
V(I)&\equiv& V(t, r)= \frac{4}{3} \pi a_{f}^3 \,a\, b\, c \,r^3 = a_f^3 \,\mathcal{V}\,,\\
A(\partial I)&\equiv& A(t, r)= 4\pi a_{f}^2 r^2\left(\frac{( a\, b)^p +(b\, c )^p +(c\, a )^p}{3} \right )^\frac{1}{p}\,, \quad \rm{with} \quad p=1.6075\,.
\end{eqnarray}
 We have used here the approximate formula for the surface area of the ellipsoid known as Thomsen's formula\cite{patial2023study}. Now, it is straightforward to write the conditions.
\begin{figure}
	\centering
	\includegraphics[scale=0.34]{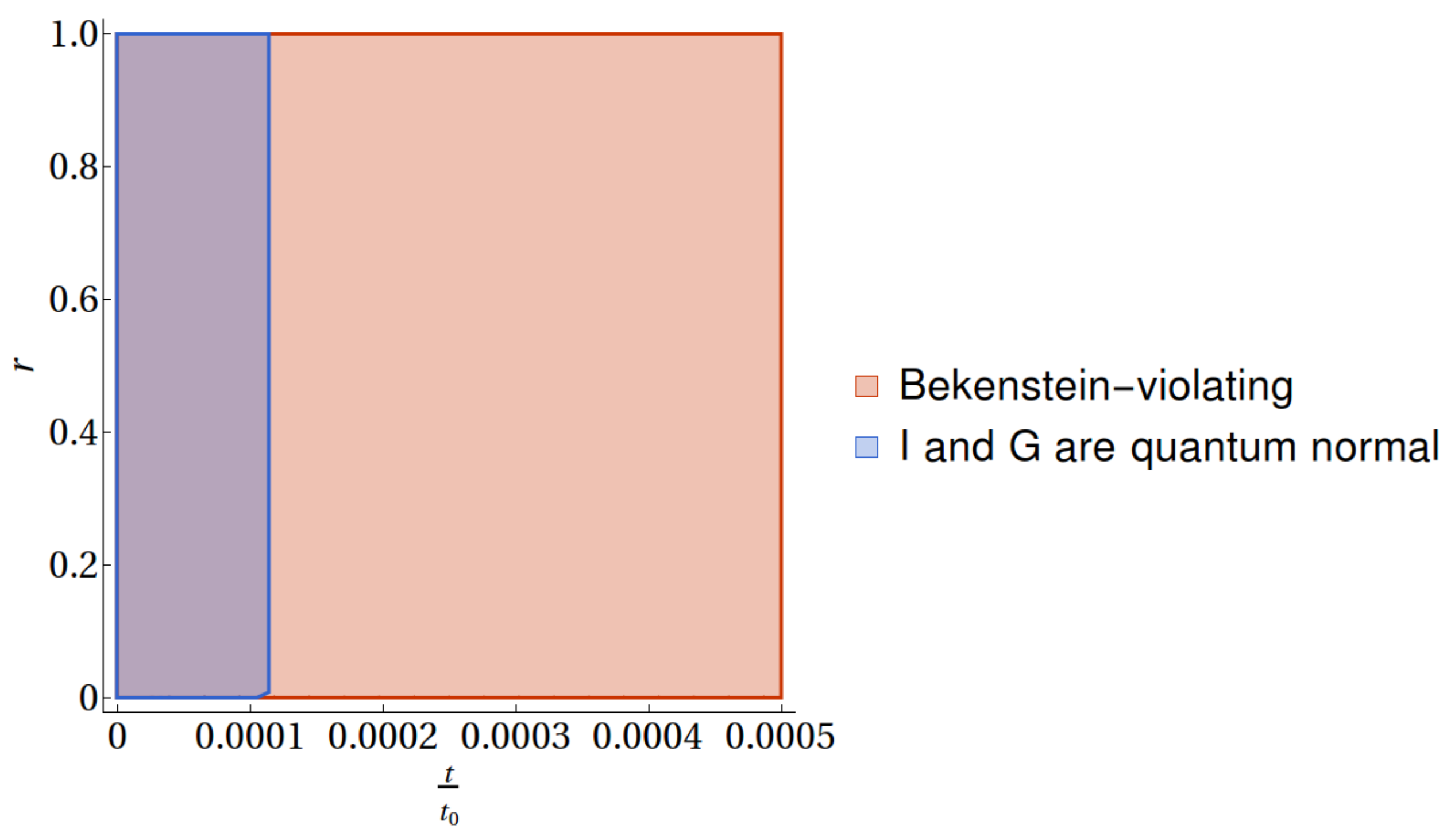}
	\caption{\em\small Plot of conditions $1$, $2$, and $3$. We set $G_N=1$ and $t_0=3*10^4$ for large values of $a$, $b$, and $c$. Condition $1$ is satisfied for all regions shown, whereas conditions $2$ and $3$ are only satisfied when $ t/t_0 \lesssim 0.0001$. The plot shows that all the conditions can be satisfied, but we are not in the semiclassical regime. (here, $t \lesssim 3<\sqrt{6\pi G_N}$) 
		\label{fig:v1}}
\end{figure}
The first condition, \Eqn{e:cond1} for this case is derived as
\begin{eqnarray}
\text{Condition 1}:\quad	\label{e:con1flrwellip}
	T_0^3> \frac{3*3^\frac{p-1}{p}\left((ab)^{p} +(bc)^{p} +(ca)^{p}\right)^\frac{1}{p} t}{16 \,a \,b\, c\, c_{\rm th}\,G_N \,r\, t_0}\,.
\end{eqnarray}
This condition\footnote{Substituting $\chi/r$ in place of $a$, $b$, and $c$ will reproduce the FLRW sphere results, where $\chi$ represents the radius of the sphere varying from $0$ to $\infty$.} can be satisfied by taking large value of $a$, $b$, and $c$, where $S_{\rm matter}(I)= s_c \mathcal{V} $, with $s_c$ being the comoving entropy.
The second \Eqn{e:sgenb1} and third \Eqn{e:sgeni} conditions are written as follows:
\begin{eqnarray}
\label{e:eqcond23}
\text{Condition 2 and 3}:  k^r\partial_r (s_c \mathcal{V})\geq \left|k^t\partial_t\left(\frac{A}{4G_N}\right)\right|+\left| k^r\partial_r\left(\frac{A}{4G_N}\right)\right|\,. 
\end{eqnarray}
Refer to Appendix \ref{a:null} for a detailed description.
After substituting volume and area, the above condition becomes
\begin{eqnarray}
\label{e:con23flrwellip}
\frac{4 c_{\rm th}T_0^3}{3} &\geq&\left| \left(\frac{(ab)^{p} +(bc)^{p} +(ca)^{p}}{3}\right)^{1/p} \frac{\sqrt{t/t_0}}{4 G_N t_0 \sqrt{a^2b^2 \cos^2\theta + c^2 \sin^2\theta (b^2\cos^2\phi+ a^2 \sin^2\phi)}}\right|  \nonumber\\
&&+\left| \left(\frac{(ab)^{p} +(bc)^{p} +(ca)^{p}}{3}\right)^{1/p}\frac{t/t_0}{2 a\,b\,c\, G_N\, r} \right|\,,
\end{eqnarray}
with $s_c=\frac{4 c_{\rm th}T_0^3}{3}$.

The second term on the right-hand side can be neglected for large values of $a$, $b$, and $c$. Then using the relation between $T_0$ and $t_0$, the above condition reduces to
\begin{eqnarray}
\label{e:tabc}
t\leq \frac{\sqrt{2 c_{\rm th}G_N}3^{3/4}(a^2b^2\cos^2\theta +c^2 \sin^2\theta(b^2\cos^2\phi +a^2\sin^2\phi  )) }{\pi^{3/2}\left((ab)^{p} +(bc)^{p} +(ca)^{p} \right)^{2/p}}\,.
\end{eqnarray}
Without loss of generality, we can take $a>b>c$, then the numerator of the right-hand side will be minimal when $\theta = \pi/2$ and $\phi=0$. \Eqn{e:tabc} becomes
\begin{eqnarray}
\label{e:tabc1}
t\leq \frac{\sqrt{2 c_{\rm th}G_N}3^{3/4}b^2c^2 }{\pi^{3/2}((ab)^{p} +(bc)^{p} +(ca)^{p} )^{2/p}} \,.
\end{eqnarray}
 However, we also require $t>>\sqrt{6\pi G_N}$ which gives us the following inequality, which can only be satisfied for a large value of $c_{\rm th}$,
\begin{eqnarray}
\label{e:tabc2}
((ab)^{8/5} +(bc)^{8/5} +(ca)^{8/5} )^{5/4}\pi^2\leq \sqrt{c_{\rm th}}3^{1/4}b^2c^2\,.
\end{eqnarray}
As illustrated in Fig. \ref{fig:v1}, both conditions \Eqn{e:con1flrwellip} and \Eqn{e:con23flrwellip} can be met, but doing so would place us outside the semiclassical regime. To remain within the semiclassical regime, a large value of $c_{\rm th}$ is required. Therefore, even in the case of an ellipsoid, the island does not exist. In the next section, we will explore an anisotropic model and demonstrate that an island can exist in such models.

As a final comment, from \eqref{e:flrweqnn}, since in FLRW model time and space factorize, and since we can pick an arbitrarily large volume, the real test of whether an island can exist or not is that the comoving entropy has to fulfil the following:
\be
s_c\geq f(t)\frac{\mathcal{A}}{\partial_r \mathcal{V}}\,,
\ee
where $f(t)$ is some time-dependent function, and $\mathcal{A},\mathcal{V}$ are the coordinate surface area and volume of the island candidate, i.e. the surface area and volume in Euclidean space. In the spherical case we got $\frac{\mathcal{A}}{\partial_r \mathcal{V}}=1$, while for the ellipsoid it turned out to be larger $\frac{\mathcal{A}}{\partial_r \mathcal{V}}>1$. Thus, it is tempting to find an optimal candidate of an island, such that $\frac{\mathcal{A}}{\partial_r \mathcal{V}}$ is minimal. If so, we can always consider whether that minimal shape is an island candidate and use it to derive no-go theorems.

\section{Bianchi type \texorpdfstring{$\rm \RomanNumeralCaps{1}$}{TEXT} Model}
\label{s:model}
The Bianchi universe models are spatially homogeneous and anisotropic, classified into nine different types and two main classes by Bianchi\cite{bianchi2001three}. 
We will be considering here the radiation-dominated Bianchi type $\rm \RomanNumeralCaps{1}$  as outlined in \cite{russell2017oscillatory,russell2014bianchi}. The metric representation for this model is given as :
\begin{eqnarray}
\label{e:metricbb}
\mathrm{d}s^2 = -\mathrm{d}t^2+a_1^2(t)\mathrm{d}x^2+ a_2^2(t) \mathrm{d}y^2 + a_3^2(t)\mathrm{d}z^2\,,
\end{eqnarray}
where $a_1$, $a_2$ and $a_3$ are the scale factors of the three-dimensional space and are functions of time. The model reduces to the flat FLRW solution if $a_1=a_2=a_3$. In general, each spatial direction has its own expansion rate $H_i\equiv \dot{a}_i/a_i$. The normalized scale factors are found as follows:
\begin{eqnarray}
\label{e:ani}
a_{nr,i}=e^{-2\alpha_{r,i}\left(\sqrt{\frac{t_0}{t}}-1 \right)}\left( \frac{t}{t_0}\right)^{1/2}\,,
\end{eqnarray}
where the exponential term is the anisotropic expansion/contraction, and $(t/t_0)^{1/2}$ is the contribution from the standard FLRW model. Here index $n$ stands for the normalization of the scale parameters to the present-day $t_0$, and index $r$ stands for the fact that the universe is filled with radiation. The dimensionless, $\alpha_i$ satisfy the following constraint:
\begin{eqnarray}
\label{e:intconst}
\sum_{i=1}^{3}\alpha_i=0\,.
\end{eqnarray}
In order to determine the turning behaviors of the scale factors, we set the directional Hubble parameter to be greater than zero.
A direct calculation gives
 \begin{eqnarray}
 \label{e:hexp}
H_i=\left(\alpha_i\sqrt{\frac{t_0}{t}}+\frac{1}{2}\right)\frac{1}{t}\Rightarrow \alpha_i>-\frac{1}{2}\sqrt{\frac{t}{t_0}}, \quad \rm{Expansion}\,;
 \end{eqnarray}
 otherwise, contraction. This behavior of switching from the previous contraction to expansion is also known as a bouncing behavior of the scale factor.
\subsection{Comoving entropy and local thermal equilibrium in Bianchi type \texorpdfstring{$\rm \RomanNumeralCaps{1}$}{TEXT} universe}
For the Bianchi type $\rm \RomanNumeralCaps{1}$ radiation case, assuming local thermal equilibrium, we also have a constant comoving entropy,
\begin{eqnarray}
\label{e:sc}
s_c=\frac{4\rho_0}{3T_0}=\frac{4c_{\rm th} T_0^3}{3}\,, \quad \quad c_{\rm th}=\frac{\pi^2}{30}g_*(T).
\end{eqnarray}
The derivation is only slightly generalized from the FLRW one.
The first and second law of thermodynamics, in the general case of a variable number of particles, has the form
\begin{eqnarray}
\label{e:firstlaw}
\mathrm{d}E=T\mathrm{d}S -p\mathrm{d}V + \sum_{i}\mu_i \mathrm{d}N_i\,,
\end{eqnarray}
where $S$ is the entropy of the system and the subscript $i$ labels particle species. Energy $E$ and the number of particles are extensive quantities proportional to the volume of the system, while temperature and pressure are local characteristics independent of the volume. Introducing densities,
\begin{eqnarray}
\label{e:den}
\rho\equiv\frac{E}{v}\,,\quad n\equiv\frac{N}{V}\,,\quad s\equiv \frac{S}{V}\,.
\end{eqnarray}
$V$ is the physical volume of the considered part of the universe at a time when the scale factor is $a$; thus, we have $V=(a_1a_2a_3)xyz=a^3xyz$. Assuming that for any physical volume $\mathrm{d}N_i=0$,
\Eqn{e:firstlaw} can be written as follows:
\begin{eqnarray}
\label{e:firstlaw1}
T\mathrm{d}S = \mathrm{d}(\rho V) + p\mathrm{d}V= (\rho+p)\mathrm{d}V +V\mathrm{d}\rho\,.
\end{eqnarray}
According to the second law of thermodynamics, the entropy of any closed system can only increase, and it stays constant for equilibrium evolution, i.e., slow evolution during which the system always remains in thermal equilibrium. Let us see for an expanding universe, assuming that the evolution of cosmic medium is close to equilibrium,
\begin{eqnarray}
\label{e:law11}
T\frac{\mathrm{d}S}{\mathrm{d}t}&=& (\rho+p)\frac{\mathrm{d}V}{\mathrm{d}t} + V\frac{\mathrm{d}\rho}{\mathrm{d}t}\,,\nonumber \\ 
&=&xyz\left[ \dot{a}_1a_2a_3 +  a_1\dot{a}_2a_3+a_1a_2\dot{a}_3  \right](\rho+p) +a_1a_2a_3xyz \dot{\rho}\,,\nonumber\\ 
&=&a_1 a_2 a_3xyz \left[ \left(\frac{\dot{a}_1}{a_1} +\frac{\dot{a}_2}{a_2}+\frac{\dot{a}_3}{a_3} \right)(\rho+p) +\dot{\rho}  \right] =0\,,
\end{eqnarray}
where we have used the covariant conservation of energy-momentum tensor in the expanding universe. We see that total entropy $S$ is conserved, which can be written as %
\begin{eqnarray}
\label{e:com}
S=sV=sa_1a_2a_3xyz=s_c xyz=\rm{constant}\,.
\end{eqnarray}
where we have defined $s_c$ as the comoving entropy and we have $\dot{s_c}=0$.
%
\subsection{``Spherical'' island}
Consider a spherical region in Bianchi type $\rm \RomanNumeralCaps{1}$ universe at $t=t_0$. We are taking all the scale factors at $t=t_0$ to be unity. 
As the time progresses, i.e., for $t\neq t_0$, the spherical region deforms according to the scale factors $a_1(t)$, $a_2(t)$, and $a_3(t)$, as defined in \Eqn{e:ani}, in the respective $x$, $y$, and $z$ direction. Thus, in other times, $t\neq t_0$, this region is not spherical but ellipsoidal. We adopt the following parametrization:
\begin{eqnarray}
    \label{e:ellippara}
x&=& \chi \sin \theta \cos\phi\,,\\
y&=& \chi \sin \theta \sin\phi\,,\\
z&=& \chi \cos \theta\,,
\end{eqnarray}
where, $\chi=\sqrt{x^2+y^2+z^2}$ is the radius of the sphere, varying from $0$ to $\infty$. On the surface, $x,y$ and $z$ depend only on $\theta$ and $\phi$. The volume and area of the sphere as functions of time $t$ and radius $\chi$ are given by
\begin{eqnarray}
\label{e:earea}
V(I)&\equiv& V(t,\theta,\phi)= \frac{4}{3} \pi a_1 a_2 a_3 \chi^3 = a_1 a_2 a_3 \mathcal{V}\,,\\
A(\partial I)&\equiv& A(t,\theta,\phi)= 4\pi \chi^2 \left(\frac{(a_1 a_2 )^p +(a_2 a_3 )^p +(a_1 a_3 )^p}{3} \right )^\frac{1}{p}\,, \quad \rm{with} \quad p=1.6075\,.
\end{eqnarray}
From \cite{ben2023islands}, the possibility of an island is if $t<<t_0$, that we shall now consider.
The first condition \Eqn{e:cond1} is written as follows:
\begin{eqnarray}
\label{e:econ1}
\text{Condition 1}:\quad 
\frac{4c_{th}T_0^3}{3}&>&\frac{3}{4G_N \chi}\frac{t}{t_0}\left(\frac{e_1^{2p}+ e_2^{2p} + e_3^{2p}}{3} \right)^{1/p}\,,
\end{eqnarray}
where we defined:
\begin{eqnarray}
    \label{e:se1}
    e_i=\exp\left(\left(\sqrt{\frac{t_0}{t}}-1\right)\alpha_i\right)\,, \quad \text{for} \quad i=1,2,3\,.
\end{eqnarray}
The above condition can always be satisfied by taking large enough $\chi$. 
\\
\text{Condition 2 and 3}: For the region I and G to be quantum normal, the inequality given below must hold:
\begin{eqnarray}
\label{e:econ2}
\frac{4 c_{\rm th}T_0^3}{3} &\geq& \Bigg| \frac{t}{2G_N \chi t_0} \Bigg\{\frac{e_1^{2p}+ e_2^{2p} + e_3^{2p}}{3}  \Bigg\}^{1/p} 
\Bigg|+ \nonumber\\
&&\Bigg| \frac{e_1^{2p}\left(\sqrt{\frac{t}{t_0}}-\alpha_1 \right) + e_2^{2p}\left(\sqrt{\frac{t}{t_0}}-\alpha_2 \right)+ e_3^{2p}\left(\sqrt{\frac{t}{t_0}}-\alpha_3 \right) }{ 3^{1/p} G_N t_0 4 \left(e_1^{2p}+e_2^{2p}+e_3^{2p} \right)^{(1-p)/p}\sqrt{ e_3^4\cos^2\theta + \sin^2\theta \left(e_1^4\cos^2\phi + e_2^4 \sin^2\phi \right)}} \Bigg|\,,
\end{eqnarray}
with
\begin{eqnarray}
\label{e:eidef}
e_i=\exp\left(\left(\sqrt{\frac{t_0}{t}}-1\right)\alpha_i\right)\,.
\end{eqnarray}
The first term in the above equation can be neglected for large $\chi$. We find the value of $\theta$ and $\phi$ such that the second term on the right-hand side is maximal. If the condition is satisfied for this maximal value, then it can be satisfied for any value of $\theta$ and $\phi$. 
Here, we are minimizing the given function below (Refer to Appendix \ref{a:alphacon}),
\begin{eqnarray}
\label{e:fthphi}
f(\theta,\phi)=e_3^4\cos^2\theta + \sin^2\theta \left(e_1^4\cos^2\phi + e_2^4 \sin^2\phi \right)\,.
\end{eqnarray}
By requiring that the determinant of the Hessian matrix of the function is greater than zero and that the second derivative of the function with respect to $\theta$ is greater than zero, we obtain the following constraints for a given value of $\theta$ and $\phi$ for which $f(\theta, \phi)$ attains the minimum value,
\begin{eqnarray}
\label{e:abccond2}
e_3>e_1 \quad \text{and} \quad e_2>e_1 \,,\quad \text{for} &\quad& f(\pi/2, 0)=f(\pi/2, \pi)=f(\pi/2, 2\pi)= e_1^4\,,\\
e_3>e_2 \quad \text{and} \quad  e_1>e_2 \,,\quad \text{for} &\quad& f(\pi/2, \pi/2)=  f(\pi/2, 3\pi/2)= e_2^4\,.
\end{eqnarray}
\Eqn{e:econ2} for large $\chi$, $\theta=\pi/2$, and $\phi=0$ reduces to
\begin{eqnarray}
\label{e:econ2.2}
\frac{4 c_{\rm th}T_0^3}{3} &\geq& 
\Bigg| \frac{e_1^{2p}\left(\sqrt{\frac{t}{t_0}}-\alpha_1 \right) + e_2^{2p}\left(\sqrt{\frac{t}{t_0}}-\alpha_2 \right)+ e_3^{2p}\left(\sqrt{\frac{t}{t_0}}-\alpha_3 \right) }{ 3^{1/p} G_N t_0 4 \left(e_1^{2p}+e_2^{2p}+e_3^{2p} \right)^{(1-p)/p}e_1^2} \Bigg|\,,
\end{eqnarray}
Given that $e_{2,3}>e_1$, it follows that $\alpha_{2,3}>\alpha_1$. The sum of three $\alpha' s$ being zero implies that $\alpha_1<0$, and $\alpha_{2,3}$ can be either positive or negative. Positive $\alpha$ implies an expansion in that particular direction. For negative $\alpha$, when $\frac{1}{2}\sqrt{\frac{t}{t_0}}$ becomes greater than $|\alpha|$, the direction switches from contraction to expansion following \Eqn{e:hexp}.\\
\begin{figure}
	\centering
	\includegraphics[scale=0.32]{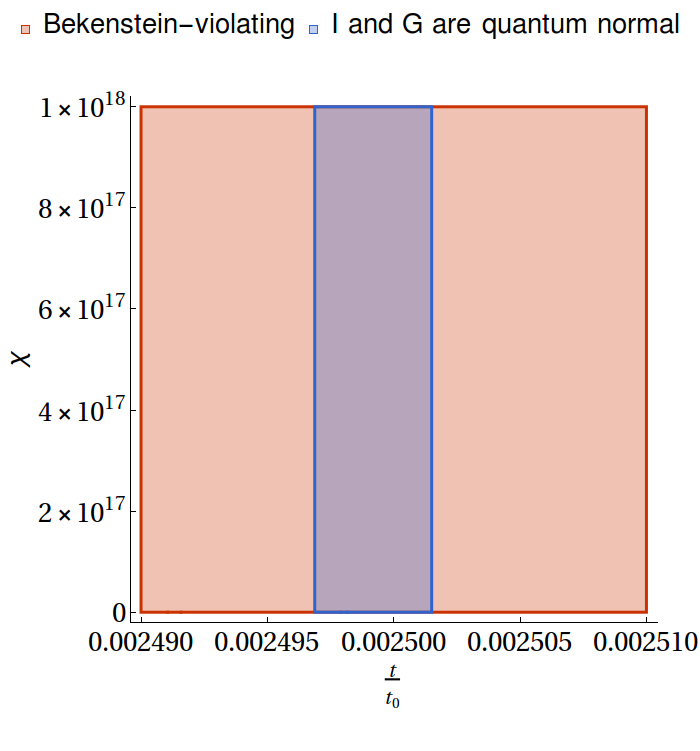}\includegraphics[scale=0.32]{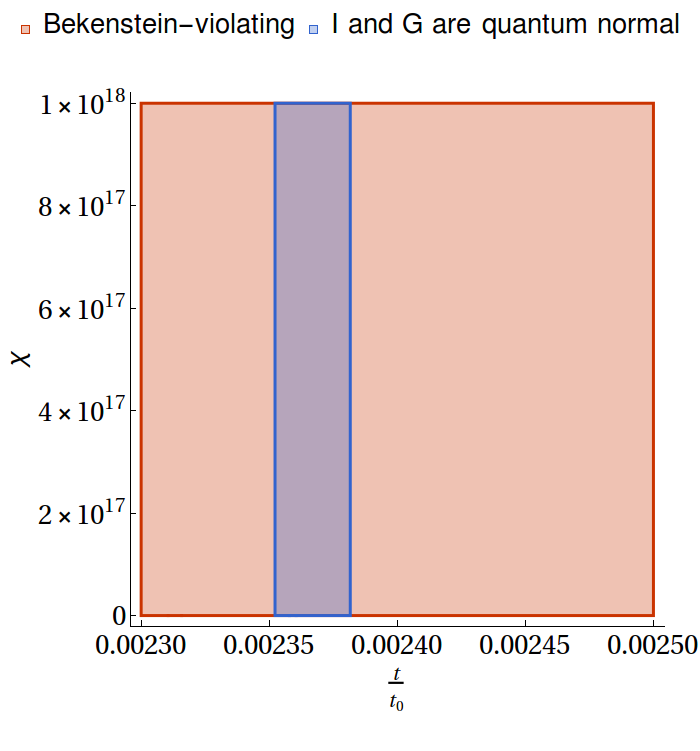}
	\caption{\em\small Plot of the region where conditions are satisfied. Case 1: $\alpha_1=-0.1$ and $\alpha_2=\alpha_3=0.05 $. Expansion in $x$ direction starts when $t/t_0$ crosses $0.04$. Case 2: $\alpha_1=-0.026$, $\alpha_2=-0.024$, and $\alpha_3=0.05$. We set $G_N=1$ and $t_0=3*10^5$ for $\theta=\pi/2$ and $\phi=0$. All the conditions are satisfied in the overlapping region.
		\label{fig:v2}}
\end{figure}
\textit{Case 1:} $\alpha_1<0$ and $\alpha_{2,3}>0$.
This implies that $e_{2,3}^p$ is large. For the above inequality to hold, the term in the bracket $\sqrt{t/t_0} - \alpha_{2,3}$ must be close to zero. Then $\alpha_1 \sim -2\sqrt{t/t_0}$. Thus, the condition for expansion, $\alpha_i >-\frac{1}{2}\sqrt{t/t_0}$ is not fulfilled. That means that, in this case, the $x$ direction will be contracting when all the conditions for the existence of the island are satisfied.\\
\textit{Case 2:} $\alpha_{1,2}<0$, $\alpha_3>0$ and $\alpha_1<\alpha_2$. \\
The first two terms in the numerator of the right-hand side of \Eqn{e:econ2.2} will be positive. We want the third term to be negative such that it cancels out the sum of the first two terms and satisfies the above condition.
Now, $\alpha_3$ can be less than, greater than or close to $\sqrt{t/t_0}$.
\begin{eqnarray}
\label{e:alpha23con}
\alpha_3<\sqrt{\frac{t}{t_0}} &\implies &  e_3^{2p}\left(\sqrt{\frac{t}{t_0}}-\alpha_3 \right)>0\,, \quad \text{inequality will not be satisfied}\,,\\
\label{e:al}
\alpha_3 \gtrsim \sqrt{\frac{t}{t_0}} &\implies  & e_3^{2p}\left(\sqrt{\frac{t}{t_0}}-\alpha_3 \right)<0\,, \quad \text{inequality can be satisfied} \,.
\end{eqnarray}
Now from \Eqn{e:al} and the condition that the sum of $\alpha$'s is zero, we can get the following constraint on the value of $\alpha_1$ and $\alpha_2$:
\begin{eqnarray}
&& \alpha_1< -\frac{1}{2}\sqrt{\frac{t}{t_0}} \quad \text{and} \quad \alpha_2>-\frac{1}{2}\sqrt{\frac{t}{t_0}}\,, \nonumber \\ && \alpha_1< -\frac{1}{2}\sqrt{\frac{t}{t_0}} \quad \text{and} \quad  \alpha_2 < -\frac{1}{2}\sqrt{\frac{t}{t_0}}\,.
\end{eqnarray}
From the above, we conclude that at least one of the directions ($x$ - direction) must be contracting to fulfil the condition. These cases are illustrated in Fig. (\ref{fig:v2}). We see that when $t<<t_0$, we found a region for which all the conditions are satisfied and an island exists, but at least one of the directions must be contracting. This region is still in the semiclassical regime, unlike the radiation-dominated FLRW model.

As our model converges to FLRW metric at $t=t_0$, we obtain similar conditions at the normalization time. This can be shown by taking $t=t_0$ in \Eqn{e:econ1},
\begin{eqnarray}
	\label{e:econ1t0}
	\frac{4c_{th}T_0^3}{3}>\frac{3}{4G_N \chi}\,,
\end{eqnarray}
and \Eqn{e:econ2},
\begin{eqnarray}
\label{e:co231}
\frac{4c_{\rm th}T_0^3}{3}\geq \left| \frac{1}{4G_N t_0}\right|+\left|\frac{1}{2 \chi G_N}\right|\,.
\end{eqnarray}
A similar analysis was conducted for an ellipsoid at $t=t_0$, following the parametrization defined in \Eqn{e:flrwellippara}. The necessary conditions and results are detailed in Appendix \ref{s:ellipbia}. The conclusions drawn were analogous to those obtained for a sphere.
\section{Summary}
\label{s:summary}
In this paper, we focus on how anisotropy affects the existence of an island. We also examine the case
of an ellipsoidal island, and as expected, the conclusions are similar to those of a spherical island. We
study the radiation-dominated flat FLRW and Bianchi type  $\rm \RomanNumeralCaps{1}$ models. For both models, we explored
the necessary conditions required for the existence of an island. We found the following:

1. Islands do not exist in the radiation-dominated FLRW Universe. Although the conditions are satisfied for $t\ll t_0$, this regime is outside the semiclassical domain. Our analysis of both spherical and ellipsoidal island cases led to similar conclusions. Unless there are additional energy scales such as cosmological constant, spatial curvature or other fluids, the only possible loophole in this case is the existence of many relativistic degrees of freedom, such that $c_{th} \gg 1$. Regarding the shape of an island in FLRW universes, the conditions include an interesting geometrical ratio $\frac{\mathcal{A}}{\partial_r\mathcal{V}}$. Finding some minimal value for this ratio will reduce island finding to a single shape---the one with the minimal ratio.

2. For the radiation-dominated anisotropic Bianchi type $\rm \RomanNumeralCaps{1}$ model, we found that islands do exist for $t\ll t_0$ in the semiclassical regime. It is around the time when the scale factors show the bouncing behavior, i.e., switching from the previous contraction to expansion.
As time progresses, Bianchi type $\rm \RomanNumeralCaps{1}$ universe approaches the FLRW universe following the isotropization criteria. And at $t=t_0$, we obtain similar results as in the case of the FLRW model. The anisotropy $K_i$ or $\alpha_i$ is again an additional energy scale on top of the temperature of radiation, allowing for the possible formation of islands. These results are valid both for spherical and ellipsoidal island candidates.

Our analysis further strengthens the idea that islands in cosmology generically occur near a bouncing behavior when some spatial direction(s) switches from contraction to expansion and that the potential shape of the island does not affect its existence.

Additionally, one can investigate the reason behind the formation of islands near the bouncing behavior in anisotropic universes. What mechanisms give rise to the large amounts of entanglement necessary for the formation of an island? Obviously, it will be connected to quantum effects, perhaps wormholes, at that time \cite{balcerzak2021entanglement,mithani2012collapse,PhysRevD.38.1761}\footnote{Note that time is ill-defined in quantum cosmology \cite{kiefer2022time}.}. In the Bianchi type $\rm \RomanNumeralCaps{1}$ model, a bounce represents a critical point because the Hubble parameter $H$ transitions from negative (during contraction) to positive (during expansion), where $(\dot{a}=0)$  momentarily vanishes. The maxima or minima of expansion, dubbed as critical points
in \cite{balcerzak2021entanglement}, exhibit high ``quantumness". This observation provides a comparative framework for validating our hypothesis. However, these works assumed two gravitating universes using third quantization, while here, we consider a gravitating system entangled with an auxiliary nongravitating one. Hence, these two different setups may have different temperatures and behaviors depending on the way the thermofield double is constructed \cite{Balasubramanian:2021wgd}.

\section*{ACKNOWLEDGMENTS}
I.B.D. and A.S. are supported in part by the ``Program
of Support of High Energy Physics'' Grant by the Israeli Council for Higher
Education.

\appendix 
\section{APPENDIX: MATHEMATICAL DERIVATIONS AND SUPPLEMENTARY DETAILS}
\label{a:appen}
\subsection{Conditions \texorpdfstring{$2$ and $3$}{TEXT} for ellipsoid in FLRW model }
\label{a:null}
The second and third conditions state that $I$ and $G$ are quantum normal regions. This is defined in \Eqn{e:sgenb1} and \Eqn{e:sgenb2}. Consider the case of an ellipsoid in the FLRW universe. On the surface of the ellipsoid we will have $r=1$ and \Eqn{e:sufofell} takes the form:
\begin{eqnarray}
\label{e:sufofell2}
f=\frac{x^2}{a^2}+\frac{y^2}{b^2}+\frac{z^2}{c^2}=1=constant\,.
\end{eqnarray}
The vector field normal to its surface is
\begin{eqnarray}
\label{e:kmu}
k^\mu=g^{\mu \nu}\nabla_\nu f\,.
\end{eqnarray}
By using the metric $g_{\mu \nu}$ as defined in \Eqn{e:flrwellipmetric} in the above equation, we get
\begin{eqnarray}
\label{e:kmu2.2}
k^1=\frac{1}{a_f^2}\frac{2 x}{a^2}\,,\quad k^2=\frac{1}{a_f^2}\frac{2 y}{b^2}\,,\quad k^3=\frac{1}{a_f^2}\frac{2 z}{c^2}\,, 
\end{eqnarray}
where $a_f$ denotes the scale factor in the FLRW universe. The subscript $f$ distinguishes it from the semimajor axis $a$ of the ellipsoid.
For it to be a null vector field, we will impose the condition that $k_\mu k^\mu=0$ and get 
 \begin{eqnarray}
 \label{e:k00}
 k^0 \propto \pm \frac{1}{a_f}\sqrt{\frac{x^2}{a^4}+\frac{y^2}{ b^4}+\frac{z^2}{c^4} }\,.
 \end{eqnarray}
We get the deformation along the null vector normal to the surface  
\begin{eqnarray}
\label{e:kmufind}
k^{\mu}\partial_\mu   &\propto& \left( \mp \frac{1}{a_f} \sqrt{\frac{x^2}{a^4}+\frac{y^2}{ b^4}+\frac{z^2}{c^4}} \partial_t + \frac{x}{a_f^2 a^2 }\partial_x + \frac{y}{a_f^2 b^2}\partial_y + \frac{z}{a_f^2 c^2}\partial_z \right)  \,. 
\end{eqnarray}
Partial derivatives for the parametrization defined in \Eqn{e:flrwellippara}: 
\begin{eqnarray}
\label{e:partia}
\partial_x&=& \frac{\cos\phi \sin\theta}{a} \partial_r + \frac{\cos\phi \cos\theta}{r}\partial_\theta - \frac{\sin\phi}{a\, r \sin \theta}\partial_\phi \,,\\
\partial_y&=&  \frac{\sin\phi \sin\theta}{b} \partial_r +\frac{\sin\phi \cos\theta}{b r}\partial_\phi + \frac{\cos\phi}{b\, r \sin \theta}\partial_\theta \,,\\
\partial_z&=& \frac{\cos\theta}{c}\partial_r -\frac{\sin\theta}{c\, r}\partial_\phi\,.
\end{eqnarray}
\Eqn{e:kmufind} then becomes
\begin{eqnarray}
\label{e:nuv173}
k^\mu \partial_\mu &\propto&\mp \frac{1}{a_f} \sqrt{ \frac{\sin^2\theta \cos^2\phi}{a^2} + \frac{\sin^2\theta \sin^2\phi}{b^2} + \frac{\cos^2\theta}{c^2}}\partial_t \nonumber \\
&& + \frac{1}{a_f^2}\left( \frac{\sin^2\theta \cos^2\phi }{a^2} + \frac{\sin^2\theta \sin^2\phi }{b^2} + \frac{\cos^2\theta}{c^2}\right)\partial_r \nonumber\\ 
&& +\frac{1}{a_f^2}\left(\frac{\cos\theta \sin\theta \cos^2\phi}{r\, a^2 }+\frac{\cos\theta \sin\theta \sin^2\phi}{r\, b^2 }-\frac{\cos\theta \sin\theta}{r\, c^2 }\right)\partial_\theta \nonumber\\ 
&&+ \frac{1}{a_f^2}\left( -\frac{\cos\phi \sin\phi}{r\, a^2 }+ \frac{\cos\phi \sin\phi}{r\, b^2 } \right)\partial_\phi\,.
\end{eqnarray}
 Conditions $2$ and $3$ are given as
\begin{eqnarray}
	\label{e:con2eq1}
	\text{Condition 2}:&& (\pm k^t\partial_t + k^i\partial_i)S_{\rm gen}(I)\geq 0 \implies k^i\partial_i (s_c \mathcal{V})\geq \pm k^t\partial_t\left(\frac{A}{4G_N}\right)- k^i\partial_i\left(\frac{A}{4G_N}\right),\\
	\text{Condition 3}:&& (\pm k^t\partial_t + k^i\partial_i)S_{\rm gen}(G)\leq 0 \implies k^i\partial_i (s_c \mathcal{V})\geq \pm k^t\partial_t\left(\frac{A}{4G_N}\right)+ k^i\partial_i\left(\frac{A}{4G_N}\right).
\end{eqnarray}
where $i=r,\theta$, and $\phi$. The area and volume of the ellipsoid do not depend on either $\theta$ and $\phi$, therefore, the above conditions are reduced to
\begin{eqnarray}
	\label{e:con2eq2}
	\text{Condition 2}:&&k^r\partial_r (s_c \mathcal{V})\geq \pm k^t\partial_t\left(\frac{A}{4G_N}\right)- k^r\partial_r\left(\frac{A}{4G_N}\right),\\
	\text{Condition 3}:&&  k^r\partial_r (s_c \mathcal{V})\geq \pm k^t\partial_t\left(\frac{A}{4G_N}\right)+ k^r\partial_r\left(\frac{A}{4G_N}\right)\,.
\end{eqnarray}
The above conditions can now be written together as follows:
\begin{eqnarray}
\label{e:con236}
k^r\partial_r (s_c \mathcal{V})\geq \left| k^t\partial_t\left(\frac{A}{4G_N}\right) \right| + \left|k^r\partial_r\left(\frac{A}{4G_N}\right)\right|\,.
\end{eqnarray}
%
\subsection{Conditions on \texorpdfstring{$\alpha_{1,2,3}$}{TEXT}}
\label{a:alphacon}
We can find the values of $\theta$ and $\phi$, which maximize the numerator using the second derivative test. If the inequality is not satisfied for the case of the maximal right-hand side, then it will not be satisfied for any value of $\theta$ and $\phi$. For the two-variable function
\begin{eqnarray}
\label{e:condf}
f(\theta,\phi)=e_3^4\cos^2\theta + \sin^2\theta \left(e_1^4\cos^2\phi + e_2^4 \sin^2\phi \right)\,,
\end{eqnarray}
The Hessian matrix and the corresponding determinant are given as 
\begin{eqnarray}
\label{e:condh}
H&=&\begin{bmatrix}
2 \cos 2 \theta \left(e_1^4 \cos^2\phi-e_3^4 
    + e_2^4 \sin^2\phi\right) & (e_2^4-e_1^4) \sin 2\theta \sin2\phi\\
(e_2^4-e_1^4) \sin 2\theta \sin2\phi & 2(e_2^4-e_1^4)\cos2\phi \sin^2\theta
\end{bmatrix}\,,\\
\rm{Det}[H]&=&(e_2^4-e_1^4) \Big[2\cos 2 \theta \cos \phi \sin^2\theta \{ e_1^4 +e_2^4 -2 e_3^4 +(e_1^4-e_2^4)\cos 2\phi  \}
\nonumber \\ &&+(e_2^4- e_1^4) \sin^2 2\theta \sin^2 2\phi
\Big].
\end{eqnarray}
By setting the first derivative of the function equal to zero, the critical points are $\theta=0,\pi/2,\pi$ and $\phi=0,\pi/2,\pi$. The value of the determinant at $\theta=0,\pi$ for any value of $\phi$ is zero; therefore, nothing can be concluded about these critical points. Now, for $\theta=\pi/2$ and $\phi\in \{0,\pi/2,\pi\}$, the second derivative and the determinant are given as
\begin{eqnarray}
\label{e:thetaphival}
f_{\theta \theta}=2(e_3^4-e_1^4)\,,&\quad & Det(\theta=\pi/2,\phi=\{0,\pi\})= 4(e_1^4-e_2^4)(e_1^4-e_3^4)\,,\\
f_{\theta \theta}=2(e_3^4-e_2^4)\,,&\quad& Det(\theta=\pi/2,\phi=\pi/2)=4(e_2^4-e_1^4)(e_2^4-e_1^4)\,.
\end{eqnarray}
To get the minimum value of $f$, we want its second derivative to be positive, given that the determinant is also positive. Implying these in the above equation, we get the conditions on the value of $a$, $b$ and $c$ and the minimum value of $f$ as follows:
\begin{eqnarray}
\label{e:abccondyy}
e_3>e_1 \quad \text{and} \quad e_2>e_1 \,\quad \text{for} &\quad& f(\pi/2, 0)=f(\pi/2, \pi)=f(\pi/2, 2\pi)= e_1^4\,,\\
e_3>e_2 \quad \text{and} \quad e_1>e_2 \,\quad \text{for} &\quad& f(\pi/2, \pi/2)=  f(\pi/2, 3\pi/2)= e_2^4\,.
\end{eqnarray}
\subsection{Ellipsoidal island candidate in Bianchi type  \texorpdfstring{$\rm \RomanNumeralCaps{1}$}{TEXT} model}
\label{s:ellipbia}
Let us perform a similar analysis for the case of an ellipsoidal island. We will use the parametrization defined in \Eqn{e:flrwellippara}. The volume and area, in this case, with the scale factors $a_1$, $a_2$ and $a_3$, are given by
\begin{eqnarray}
\label{e:volarea}
V(I)&=& \frac{4}{3} \pi a_1 a_2 a_3 a b cr^3 = a_1 a_2 a_3 \mathcal{V}(r)\,,\\
A(\partial I)&=& 4\pi \left(\frac{( a_1 a_2 a b r^2)^p +(a_2 a_3 b c  r^2)^p +(a_1 a_3 a c r^2)^p}{3} \right )^\frac{1}{p}\,,\,\,\rm{with} \quad p=1.6075\,.
\end{eqnarray}
The conditions necessary for the existence of an island are given as follows:\\
\textit{Condition $1$:}
\begin{eqnarray}
\label{e:bielc1}
\frac{4c_{\rm th} T_0^3}{3}&>&\frac{9}{4\,a\,b\,c\,G_N\,r}\frac{t}{t_0}\left(\frac{(ac)^{8/5}e_2^{16/5}+(bc)^{8/5}e_1^{16/5}+(ab)^{8/5}e_3^{16/5}}{3}\right)^{5/8}\,.
\end{eqnarray}
We can take large values for $a$, $b$, and $c$ to satisfy the above condition.
Using \Eqn{e:eqcond23} and \Eqn{e:volarea}, we get\\
\textit{Condition $2$ and $3$:}
\begin{eqnarray}
\label{e:bielc2}
\frac{4c_{\rm th} T_0^3}{3}&\geq& \left|\left(\frac{(ac)^{8/5}e_2^{16/5}+(bc)^{8/5}e_1^{16/5}+(ab)^{8/5}e_3^{16/5}}{3}\right)^{5/8}\frac{t}{t_0}\frac{1}{2\,a\,b\,c\,G_N\,r} \right|\nonumber\\
&&+\Bigg|\frac{(ac)^{8/5}\,e_2^{16/5}\left( \sqrt{\frac{t}{t_0}}-\alpha_2\right)+(ab)^{8/5}\,e_3^{16/5}\left( \sqrt{\frac{t}{t_0}}-\alpha_3\right)+(bc)^{8/5}\,e_1^{16/5}\left( \sqrt{\frac{t}{t_0}}-\alpha_1\right) }{4 G_N t_0 3^{5/8}\left( (ac)^{8/5}\,e_2^{16/5}+(ab)^{8/5}\,e_3^{16/5}+(bc)^{8/5}\,e_1^{16/5} \right)^{3/8}}\nonumber \\
&&\times
\frac{ 1 }{\sqrt{ a^2 b^2 e_3^4\cos^2\theta +c^2 \sin^2\theta\left(b^2 e_1^4\cos^2\phi +a^2 e_2^4\sin^2\phi \right) }  } \Bigg|\,.
\end{eqnarray}
Again, the first term on the rhs of the above equation can be neglected for large values of $a$, $b$, and $c$. For $a>b>c$ and $e_1<e_{2,3}$, $\theta=\pi/2 $ and $\phi=0$ gives the minimal value of the square root term in the denominator. The condition is similar to the one obtained for the sphere, and we can do a similar analysis as done before. We will obtain an island for $t<<t_0$ given that at least one of the directions is contracting. 
\bibliographystyle{JHEP}
\bibliography{bianchi.bib}

\end{document}